\documentclass[twocolumn,prl,showpacs,
amsmath,amssymb]{revtex4}
\usepackage{hyperref}
\usepackage{graphicx}
\usepackage{dcolumn}
\usepackage{bm}
\begin{document}
\title{Slow flows of yield stress fluids:\\ complex spatio-temporal behaviour
within a simple elasto-plastic model}
\author{Guillemette Picard $^{1}$, Armand Ajdari $^{1}$, Fran\c cois Lequeux $^{2}$, Lyd\'eric Bocquet $^{3}$
}
\email{lbocquet@lpmcn.univ-lyon1.fr}
\affiliation{
$^{1}$ Laboratoire P.C.T., UMR CNRS 7083, ESPCI,  10 rue Vauquelin, 75005 Paris, France\\
$^{2}$ Laboratoire P.C.M., UMR CNRS 7615, ESPCI,  10 rue Vauquelin, 75005 Paris, France\\
$^{3}$ Laboratoire P.M.C.N., UMR CNRS 5586, Universit\'e Lyon I, 69622 Villeurbanne, France\\
}
\date{\today}
\begin{abstract}
A minimal athermal model for the flow of dense disordered materials is proposed,
based on two generic ingredients: local
plastic events occuring above a microscopic yield stress, and the non-local elastic
release of the stress these events induce
in the material. A complex spatio-temporal rheological behaviour
results, with features in line with recent experimental observations.
At low shear rates, macroscopic flow actually originates from
collective correlated bursts of plastic events, taking place in dynamically generated
fragile zones.
The related correlation length
diverges algebraically at small shear rates. In confined geometries bursts 
occur preferentially close to the walls yielding an 
intermittent form of flow 
localization.
\end{abstract}
\pacs{83.50.-v, 81.40.Lm,62.20.Fe}
\maketitle
\narrowtext
Disordered dense systems often exhibit a peculiar flowing
behaviour which strongly departs from the academic Newtonian
description. A non linear rheology, with a shear rate dependence
of the viscosity, or the existence of a yield stress are typical
features of such systems. Moreover, it has been recognized
recently that these {\it global} characteristics are, in most
cases, associated with a peculiar {\it spatial} behaviour, in the
form of heterogeneous flow behaviour, where a frozen region
coexists with a flowing one (the so-called ''shear band''). A
striking remark is that such generic behaviours are observed in a
wide class of experimental systems, with very different
length/time/interaction scales, such as foams
\cite{debregeas,dennin04}, granular systems \cite{losert,dacruz},
emulsions \cite{dacruz,cous,salmon}, colloidal glasses \cite{pignon}, polymers, 
but also in simulations of granular systems, foams and
model glasses \cite{bulatov1,falk,varnik,kabla}. 
These generic features
suggest an underlying common scenario for the flow
properties,
and has motivated various macroscopic phenomenological approaches (see eg refs.
cited in \cite{cous} and \cite{varnik}).
However a consistent framework linking the global
rheology to the local microscopic dynamics is still lacking,
although some progress in this direction has been made 
in recent years \cite{barrat, lemaitre02, baret}. In
particular studies have put forward the role of local
plastic rearrangements in the global flow behaviour
\cite{langer1,falk,lemaitre04}. Such an idea actually goes back to
the Princen model for the deformation of foams \cite{Princen}~: flow occurs via a succession of
reversible elastic deformations and irreversible plastic events
(``T1'' events in foams), associated with the existence of a local
yield stress. However, if the corresponding physical picture seems
{\it a priori} quite clear, a gap still persists between this
simple microscopic scenario and 
the complex spatio-temporal organization responsible
for the rheology of these materials at finite shear rates. 

In this letter, we propose a simple mesoscopic model, constructed
on the basis of two {\it minimal} and {\it generic} ingredients :
localized plastic events associated with a
microscopic yield stress, and the
resulting elastic relaxation of the stress over
the system. 
We then show that the simplicity of the description
contrasts with the complex rheological behaviour deriving from it.
In particular we find the global rheology to be associated with a complex
spatio-temporal organization which builds up as the system is
sheared steadily, with an intermittent behaviour
corresponding to "bursts" of correlated events, the typical size 
of which diverges at small shear rate. 
We argue that in its present simple form our model seems to capture
many observed experimental features and thus stands as
a promising starting point
for the elaboration
of a generic scenario for the slow flow of yield stress fluids.

Let us now precise the ingredients of our approach, which we implement here
in the simplified frame of a 2D scalar approach, focusing only on the simple shear components
of the stress and strain.
We consider a two dimensional material to which a global shear rate $\dot\gamma$
is applied (corresponding to a $z$ dependent displacement in the x direction).
The material is described at a coarse grained level, intermediate between
the microscopic (particle) and macroscopic scale. The quantity of
interest is the xz component of the time dependent local shear stress $\sigma(x,z;t)$.
First, without entering into details at this level, a few basic rules are stated : (i)
below a (locally defined) yield stress $\sigma_Y$, the system responds elastically
to the imposed deformation; (ii) above $\sigma_Y$, plastic events may occur in the system
(along laws discussed in the following); (iii) plastic events take the form
of a localized shear strain; (iv) such a plastic
event induces a long range {\it elastic} perturbation of the shear stress field in the material.
A few remarks can be done at this level. First, although the notion of individual
events is quite intuitive, in particular in foams, it has been evidenced unambiguously only recently
at the microscopic level in disordered systems \cite{falk,lemaitre04}.
Second, the shear stress perturbation alluded to in (iv) is computed exactly 
within the framework of tensorial linear elasticity for an isotropic incompressible material
as reported in Ref. \cite{guil04}. This provides the explicit Green function, $G_{xzxz}$,
relating the stress variation, $\delta\sigma$, at any point in the system, to the xz
component of the plastic strain
$\epsilon^{pl}(\{x',z'\};t)$, associated  with the plastic event localized at $\{x',z'\}$.
Using the simpler notation $G$ for this function yields
\begin{equation}
\delta\sigma(\{x,z\};t)=2 \mu \int d{\bf r}'\ G(x,x',z,z') \epsilon^{pl}(\{x',z'\};t)
\label{stress1}
\end{equation}
The shear modulus $\mu$ has been exhibited for convenience. 
In a 2D infinite system, $G$ decreases as $G(r)=1/\pi r^2 \cos (4\theta)$ 
- in cylindrical coordinates \{$r,\theta$\} - (in agreement with refs \cite{eshelby,lemaitre04}).
In general, its precise form  depends on the specific geometry of the system :
infinite, periodic or confined between two rigid walls \cite{guil04}.
Summing up at this point, the evolution of the shear stress field
results from the global elastic loading $\dot\gamma$
plus the perturbations induced by the localized plastic events:
\begin{equation}
\partial_t \sigma (\{x,z\},t) = \mu \dot\gamma + 2 \mu \int d{\bf r}' \ G(x,x',z,z') \dot\epsilon^{pl}(\{x',z'\};t)
\label{stress2}
\end{equation}
The last part of the modelization is the choice of a dynamical law for the plastic events,
 {\it  i.e.} the feed-back law relating the plastic relaxation $\epsilon^{pl}(\{x,z\};t)$ to
the stress field $\sigma(\{x',z'\};t'<t)$.
As in the Princen model, we choose a {\em local} relation with a threshold stress value $\sigma_Y$.
In addition,
an intrinsic time scale $\tau$ is introduced to describe the dynamics of the event.
We anticipate that this will lead to a shear rate dependence of the
dynamical structure in the flow, 
driving the system away from the critical
quasi-static limit (self-organized criticality in a related quasi-static model
was reported in \cite{Chen}) 
to a more homogenous situation at large shear rate. 
Another important outcome is that 
the local stress may exceed the yield stress $\sigma_Y$ for a finite time interval 
so that the averaged stress can also grow beyond this value, 
as observed experimentally.  

There are actually many possibilities to introduce such an
intrinsic time scale for plastic events, and few guides as to how should do so. 
We make here a simple arbitrary choice and 
assume that the system locally alternates between a purely elastic
state and a plastic state (during which stress is released), with
{\em finite transition rates}~:
$\tau_{plast}^{-1}$ is the rate of
transitions from {\it elastic to plastic}, while the reverse transition is
characterized by a time $\tau_{elast}$. Since plastic events only occur above
the yield stress $\sigma_Y$, we take $\tau_{plast}(\sigma)=\infty$ if locally
$\sigma < \sigma_Y$. We otherwise assume 
for sake of simplicity fixed values for the
$\tau_{elast}$ and $\tau_{plast}$, independent of the local stress.
In order to finalize our model, we
eventually have to quantify the amount of plastic strain released in an event
and simply assume
a Maxwell, visco-elastic like relaxation of the material in the plastic state
$\dot\epsilon_{plast}=
{1\over{2 \mu \tau}} \sigma$, with $\tau$ a mechanical relaxation time.
All the previous discussion is best summarized by
introducting a ``state variable'' $n(x,z)$ such that $n=0/1$ identifies the
elastic/plastic state~:
\begin{eqnarray}
\dot\epsilon ^{pl} (\{x,z\},t) &= \frac{1}{2 \mu \tau} n(\{x,z\},t) &\sigma(\{x,z\},t)  \nonumber \\
n(\{x,z\},t) : &  0 {\displaystyle{\mathop{\longrightarrow}^{\tau_{plast}^{-1}}_{{\rm if~ } \sigma >\sigma_Y}}} 1 &
0 {\displaystyle{\mathop{\longleftarrow}^{\tau_{elast}^{-1}}_{\forall \sigma}}} 1
\label{summary}
\end{eqnarray}
Equations (\ref{summary}) and (\ref{stress2}) constitutes our minimal starting point to describe
the dynamics of yield stress materials under flow.
Note that, as in the somewhat related analysis of Langer \cite{langer1}, neither the stress nor the state variable are convected
by the displacement field within the present simplified model.

Before turning to their resolution,
eqs (\ref{stress2}-\ref{summary}) are made dimensionless using  $\sigma_Y$ and $\tau$ as
stress and time units. An important point emerging from this procedure is 
that {\it the shear rate only appears in the form of 
the ratio} $\dot\gamma/\dot\gamma_c$, with $\dot\gamma_c=\sigma_Y/\mu\tau$. 
In this dimensionless form, our model therefore therefore points out to a very general scenario,
in which specific microscopic details are embeded in the precise values of $\sigma_Y$ and 
$\dot\gamma_c$, as already suggested by some experiments \cite{cloitre,cous}.

The dynamical equations in this dimensionless form have been solved numerically, by 
discretizing the material into blocks of elementary size $a$. 
A pseudospectral method is used, which
allows us to express easily the stress increments in reciprocal space at each timestep. 
On the other hand, the state variable $n({\rm i}~a,{\rm j}~a)$ in the block \{${\rm i,j}$\} evolves 
in real space according to the
stochastic laws enonced above. 
We have focused on two geometries of $N=(L/a)^2$ blocks~: a biperiodic geometry and a confined one where
the system is bounded by two rigid parallel walls. Practically we have chosen 
$\tau_{plast}=\tau_{elast}=\tau$ for the results reported here. 

\begin{figure}[t]
\includegraphics[width=7cm]{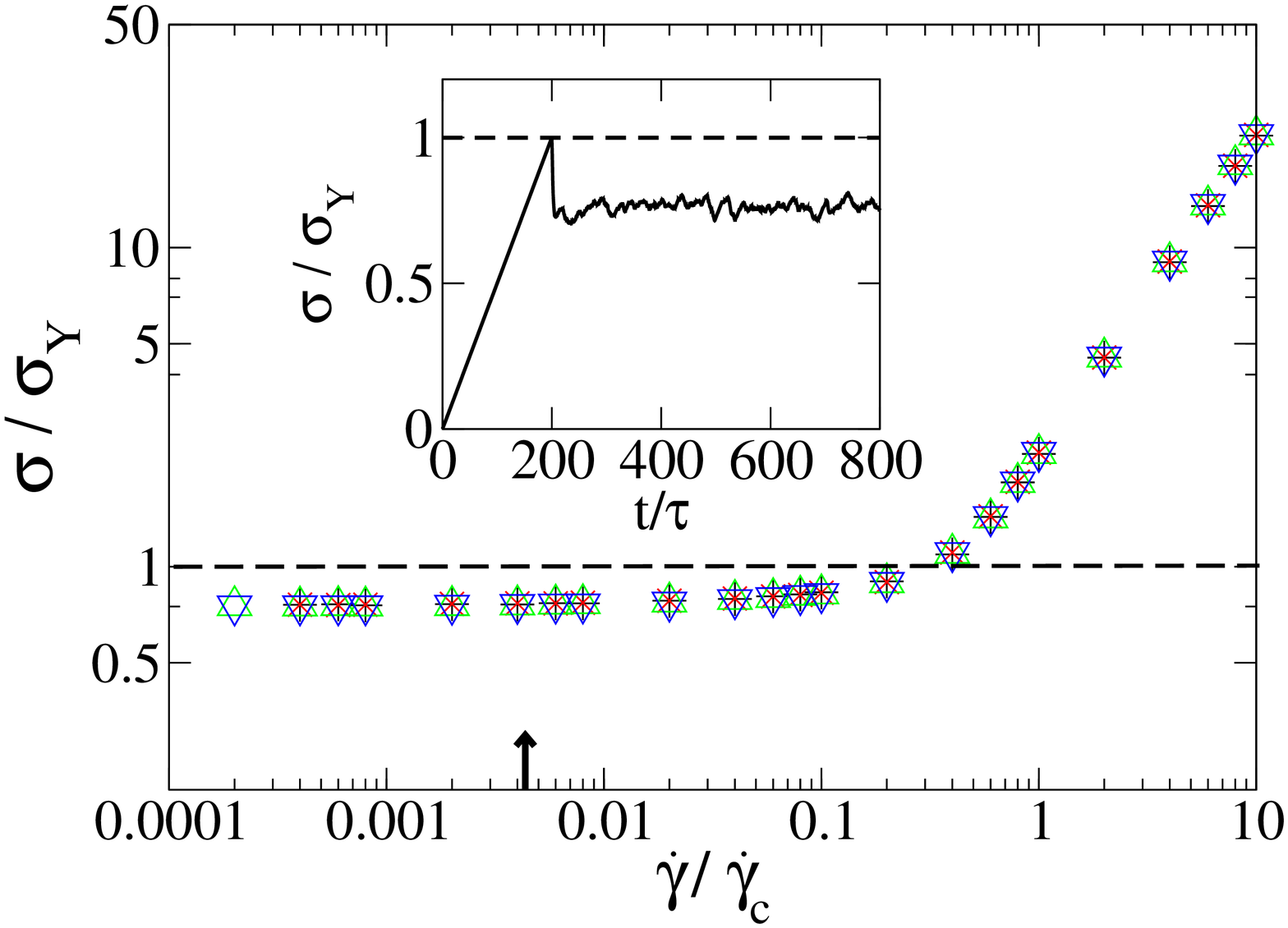}
\caption{Shear stress versus shear rate  
 in units of $\sigma_Y$ and $\dot\gamma_c=\sigma_Y/\mu\tau$ (log-log plot).
The various symbols correspond to four different
system sizes (from $4 \times 4$ to $32\times 32$ blocks). Inset: 
time dependent stress at the low shear rate corresponding to the arrow.
The dashed line corresponds to the microscopic yield stress $\sigma_Y$.
}
\label{fig1}
\end{figure}

\begin{figure}[htb]
\begin{center}
\includegraphics[width=8cm,height=3.5cm]{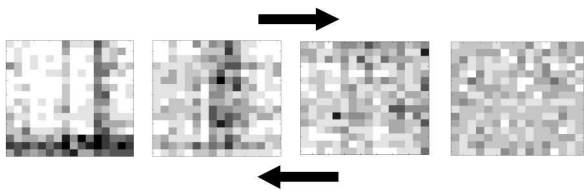}
\caption{Spatial distribution in \{x,z\} plane 
of the cumulated plastic activity for a $N=16\times16$ system after
$2N=512$ plastic events. From left to right the dimensionless shear rate, $\dot\gamma/\dot\gamma_c$, is : $5.10^{-4} ,  5.10^{-3}, 0.05, 2$.  Gray levels correspond to
the number of plastic events. Arrows indicate the imposed shear direction.}
\label{fig3}
\end{center}
\end{figure}

\begin{figure}[htb]
\includegraphics[width=7cm]{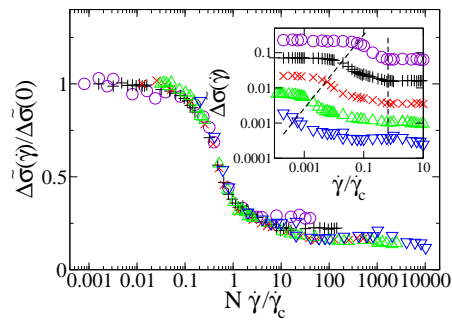}
\caption{Mean stress drop (normalized by the average stress), $\Delta\tilde\sigma$, as a function of the renormalized 
dimensionless shear rate, $N \dot\gamma/\dot\gamma_c$.
Results in the inset are shown
for five different sizes (from $2 \times 2$ to $32\times 32$ blocks from top to bottom), while in the main graph all curves are rescaled using variables
\{$N\dot\gamma/\dot\gamma_c,  \Delta\tilde\sigma(\dot\gamma)/\Delta\tilde\sigma(0)$\} to emphasize the 
scaling of the
transition between the small shear rate plateau and the intermediate regime.
The dotted lines in the inset sketch the separation between the three dynamical regimes.}
\label{fig4}
\end{figure}

\begin{figure}[htb]
\includegraphics[width=6cm]{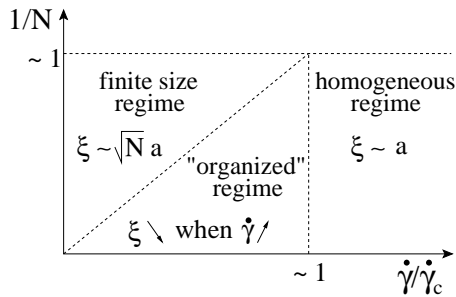}
\caption{Sketch of the emerging flow scenario in the $(\dot\gamma/\dot\gamma_c,1/N)$ plane.
Successive transitions from a homogeneous flow to an organized 
and a finite size regime occur as the correlation length $\xi$
grows from the block size to the system size, as the shear rate decreases.}
\label{fig5}
\end{figure}

We first quote the results for the biperiodic system. 
In Fig. \ref{fig1}, we plot the results for the macroscopic flow curve, which displays 
the essential features observed in experiments.
First a plateau is found at small
shear rate, defining a {\it macroscopic yield stress}
at vanishing shear rates. The latter is found to be lower than the {\it microscopic} yield stress $\sigma_Y$,
and also lower than the related peak value of the stress versus time at small shear
rates (see inset in Fig. \ref{fig1}). A different regime is found at large shear rates,
where a Newtonian behaviour is recovered.
The dynamics of the time dependent shear stress is also quite different in these two
regimes. In particular, relative stress fluctuations around the mean value increase
as the shear rate decreases (not shown), 
in agreement with observations in experiments and simulations \cite{pignon,varnik}. A zoom
on the dynamics at shorter time scale actually shows that at small shear rate
the stress exhibits successive periods of elastic raise and abrupt drops, as observed in 
the quasistatic limit in various systems \cite{lemaitre04}.
These drops encompass many events, constituting "bursts" of correlated plastic activity.  
More interestingly these
dynamically correlated events are also highly correlated in space, as emphasized in
Fig. \ref{fig3} where the spatial distribution of the cumulated plastic activity is plotted
for a given succession of plastic events. This figure clearly shows that while at high
shear rate plastic events are spatially decorrelated, a correlation pattern shows up
as the shear rate is decreased, leading to the development of  long-lived
``fragile'' zones  in the system where nearly all the plastic activity takes place.
This graph therefore suggests the development of a shear rate dependent length scale in the
system, which grows at small shear rates. In order to get more insight into this aspect,
a possible route is to measure the length via measurements of correlation function.
This is however a difficult task in general \cite{berthier} and we have followed a
different strategy here, 
analogous to finite size scaling. Namely, since such a length is associated with the
correlation of plastic events during a macroscopic stress drop, it should
show up in the statistics of stress drops. To this end, we have computed the
mean stress drop in plastic events, $\Delta\sigma$, as a function of shear rate \cite{identify}.
Results are shown in Fig. \ref{fig4} for various system
sizes. Let us first discuss the inset which exhibits the bare results for the average stress drop
normalized by the average stress, $\Delta\tilde\sigma=\Delta\sigma(\dot\gamma)/\sigma(\dot\gamma)$. 
Three different regimes can be identified for all system sizes : two plateaus at large
and small shear rates and an intermediate regime relating these two. 
We remark that the transition between the intermediate and ``saturation'' regime at low $\dot\gamma$
shifts to lower shear rates when the size of the system is increased. The system size
dependence of this transition {\it at low shear rate} is best evidenced using a rescaling procedure,
in which all different curves are found to coincide in the variables \{$N\dot\gamma/\dot\gamma_c,
\Delta\tilde\sigma(\dot\gamma)/\Delta\tilde\sigma(0)$\}. 
This rescaling in the $\dot\gamma$ variable
actually allows us to quantify the correlation spatial effect observed above. Assuming the existence
of a shear rate dependent correlation length $\xi(\dot\gamma)$ in the system,
a saturation effect is expected for the mean stress drop when $\xi(\dot\gamma)$ reaches
the system size, $N^{1/2}a$. The rescaled graph indicates that such a saturation
occurs for a fixed value of $N \dot\gamma/\dot\gamma_c$  
which suggests $\xi(\dot\gamma)\sim\dot\gamma^{-\alpha}$, with
$\alpha\simeq1/2$ from these data. 
Our model therefore explicitly yields indication
of (at least) one {\it diverging length scale at small shear rates},
a feature to our knowledge absent in previous studies of yield stress fluids.
Interestingly, the transition between the intermediate to the large shear rate regime
on Fig. \ref{fig4} (right dotted line in the inset) occurs roughly at the
characteristic shear rate  
 $\dot\gamma_c$, independent of system size, as for the macroscopic flow curve
in Fig. \ref{fig1}.
From these first results, our model yields a flow behaviour
with three different regimes as sketched
on Fig. \ref{fig5}:
(i) for $\dot\gamma> \dot\gamma_c$ (or $\sigma>\sigma_Y$),
the blocks uncorrelated in their dynamics and the flow is homogeneous ;
(ii) for $\dot\gamma<\dot\gamma_c$, correlations increase
up to a correlation length $\xi(\dot\gamma)$ which diverges algebraically at small shear rates;
(iii) at very low shear rates, the correlation length saturates at the size of the system, 
leading to a quasi-static dynamical behaviour.

We have also studied a confined geometry
where two rigid walls bound the system in the $z$ direction. A delicate technical point
is then the calculation of the Green function,
which shows that shear stress perturbation
is amplified close to the walls \cite{guil04}.
Essentially, the picture in the confined geometry is very similar to that of the biperiodic system
described above (Figs 1, 3, 4).
One important specific feature however
concerns the localization of the flow : while at high shear rate the flow is homogeneous,
at low shear rates the plastic bursts occur preferentially close to the walls.
In this last regime, the {\em average} flow corresponds to an increased shear rate close to the walls,
but this "localization on average" of the flow is only part of a complex spatio-temporal pattern.
A more detailed analysis of this regime is left for a future publication.

To sum up, we have proposed an athermal elasto-plastic model for the flow of
yield stress systems, constructed on the basis of two generic
ingredients: localized plastic events, occurring
above a microscopic yield stress with a finite duration,
and an otherwise elastic behaviour of the material (including
redistribution of stress during the events). These two ingredients
lead to a complex spatio-temporal behaviour of the system at small
shear rates. More precisely a correlation length is exhibited
which diverges at small shear rates, corresponding to intermittent
collective events (correlated bursts of plastic events), leading
to the creation of (long live) fragile zones where  the
deformation of the system takes place. These bursts take place
preferentially close to the walls. At high shear rates, this
correlation length is comparable to the size of the individual
elements which flow independently from one another.
These features are essentially compatible with recent observations
in experimental or numerical systems :
localization of the time-averaged deformation \cite{cous,debregeas,dacruz,dennin04,losert,varnik,kabla},
intermittency at low shear rate \cite{salmon,losert,varnik,lemaitre04},
a diverging length scale at small shear rate in granular systems \cite{GDR}. Moreover
numerical simulation of glassy systems \cite{varnik} show that flow heterogeneities
occur for global shear rates such that $\sigma<\sigma_Y$, a conclusion which
is recovered within our minimal model.
Although our model should be refined to take into account convection
and the full tensorial nature of the problem, 
the present early results suggest that the generic behaviors observed in the experiments and molecular
simulations originate in a minimal number of ingredients. This opens the possibility
for a coherent and robust scenario for the slow flow behaviour of disordered materials.


\begin{references}
\bibitem{identify} Stress drops can be identified using a simple criterion, checking that the stress 
at a time $t+\delta t$ is larger than the stress at time $t$. We chose $\delta t=0.01\tau$ and checked
that results are robust with respect to variations of $\delta t$.
\end{references}
\end{document}